\documentclass[a4paper,11pt]{article}
\usepackage{pos}

\title{Measurements of $C\!P$ violation with charm hadrons at Belle and Belle II}

\author*[a]{Jaeyoung Kim}
\onbehalf{on behalf of the Belle II Collaboration}

\affiliation[a]{Department of Physics, Yonsei University,\\
  03722, Seoul, Republic of Korea}

\emailAdd{jaeyoung\_kim@yonsei.ac.kr}

\abstract{Recent searches for $C\!P$ violation in charm-hadron decays at Belle and Belle II are presented.
The measurements cover $D^+_{(s)}\to\eta h^+$, $D\to\pi\pi$, $D^0\to K_S^0K_S^0$, and $D^0\to\pi^+\pi^-\pi^0$, together with the singly Cabibbo-suppressed baryon decays $\Xi_c^+\to\Sigma^+h^+h^-$ and $\Lambda_c^+\to ph^+h^-$, where $h=\pi,K$.
The Belle II analyses use the Run~1 data sample of $428\,\mathrm{fb}^{-1}$, supplemented by $980\,\mathrm{fb}^{-1}$ of Belle data for the $K_S^0K_S^0$ measurements.
We report preliminary results for the $\eta h^+$ modes, complementary tests of the isospin structure of $D\to\pi\pi$, and an application of inclusive charm-flavor tagging.
The first individual decay-rate asymmetries in these three-body charmed-baryon modes also provide tests of $U$-spin symmetry.
All reported asymmetries are consistent with $C\!P$ symmetry within the present precision.}

\FullConference{43rd International Conference on High Energy Physics (ICHEP 2026)\\
30 July -- 5 August 2026\\
Natal, Brazil\\}

\begin{document}
\maketitle

\section{Introduction}
Charm decays probe $C\!P$ violation in the up-type quark sector.
Small weak-phase differences suppress Standard Model (SM) asymmetries, while strong interactions complicate predictions.
Complementary final states and symmetry tests help clarify the origin of charm $C\!P$ violation, first observed by LHCb through
\begin{equation}
\Delta A_{C\!P}\equiv A_{C\!P}(D^0\to K^+K^-)-A_{C\!P}(D^0\to\pi^+\pi^-)
=(-15.4\pm2.9)\times10^{-4},
\end{equation}
with $5.3\sigma$ significance~\cite{lhcb2019}, followed by $3.8\sigma$ evidence for direct $C\!P$ violation in $D^0\to\pi^+\pi^-$~\cite{lhcb2023}.

Belle and Belle II~\cite{belle,belleii} collect $e^+e^-$ collision data at or near $\sqrt{s}=10.58\,\mathrm{GeV}$, corresponding to the $\Upsilon(4S)$ resonance.
Continuum $c\bar c$ production provides abundant prompt charm, and efficient neutral-particle reconstruction enables studies of $\pi^0$, $\eta$, and $K_S^0$ final states.
These Belle II results use $428\,\mathrm{fb}^{-1}$ from Run~1; the $K_S^0K_S^0$ analyses additionally use $980\,\mathrm{fb}^{-1}$ from Belle.
Charge-conjugate decays are implied.
The first uncertainty is statistical and the second systematic unless stated otherwise.

\section{\texorpdfstring{Search for $C\!P$ violation in charm hadrons}{Search for CP violation in charm hadrons}}
The decay-rate asymmetry for a charm hadron $X_c$ and final state $f$ is
\begin{equation}
A_{C\!P}(X_c\to f)=
\frac{\Gamma(X_c\to f)-\Gamma(\bar X_c\to\bar f)}
{\Gamma(X_c\to f)+\Gamma(\bar X_c\to\bar f)}.
\end{equation}
Neutral-meson asymmetries are time integrated and can contain direct and indirect contributions. Fits determine the raw yield asymmetry, which to first order is
\begin{equation}
A_{\rm raw}\simeq A_{C\!P}+A_{\rm prod}+A_{\rm det},
\end{equation}
where the additional terms describe production and detection effects.
The production asymmetry is odd as a function of $\cos\theta_{\rm CM}$, where $\theta_{\rm CM}$ is the polar angle of the charm hadron in the center-of-mass frame.
It can be canceled either by forward--backward averaging or through the use of a suitable control mode.
Kinematically matched control modes correct detection asymmetries, with any known physical $C\!P$ asymmetry of the control decay taken into account.

Charged hadrons are self-tagging; neutral $D$ flavor is conventionally identified by the slow pion in $D^{*+}\to D^0\pi^+$.
The inclusive charm-flavor tagger (CFT)~\cite{cft} instead uses charged-particle correlations in the rest of the event, achieving an effective tagging efficiency of $(47.91\pm0.07\pm0.51)\%$.

\subsection{Charm mesons}
\subsubsection{\texorpdfstring{$D^+_{(s)}\to \eta(\pi^+,K^+)$}{D(s)+ -> eta(pi+,K+)}}
Belle II presents preliminary measurements of $D^+_{(s)}\to\eta h^+$, with $h=\pi,K$. The $D^+\to\eta\pi^+$ and $D_s^+\to\eta K^+$ modes are singly Cabibbo suppressed (SCS) and sensitive to interfering amplitudes with different weak and strong phases. The $D_s^+\to\eta\pi^+$ mode is Cabibbo-favored (CF), and $D^+\to\eta K^+$ is doubly Cabibbo-suppressed.

\begin{figure}[htbp]
\centering
\begin{minipage}{0.46\linewidth}
  \centering
  \includegraphics[width=\linewidth]{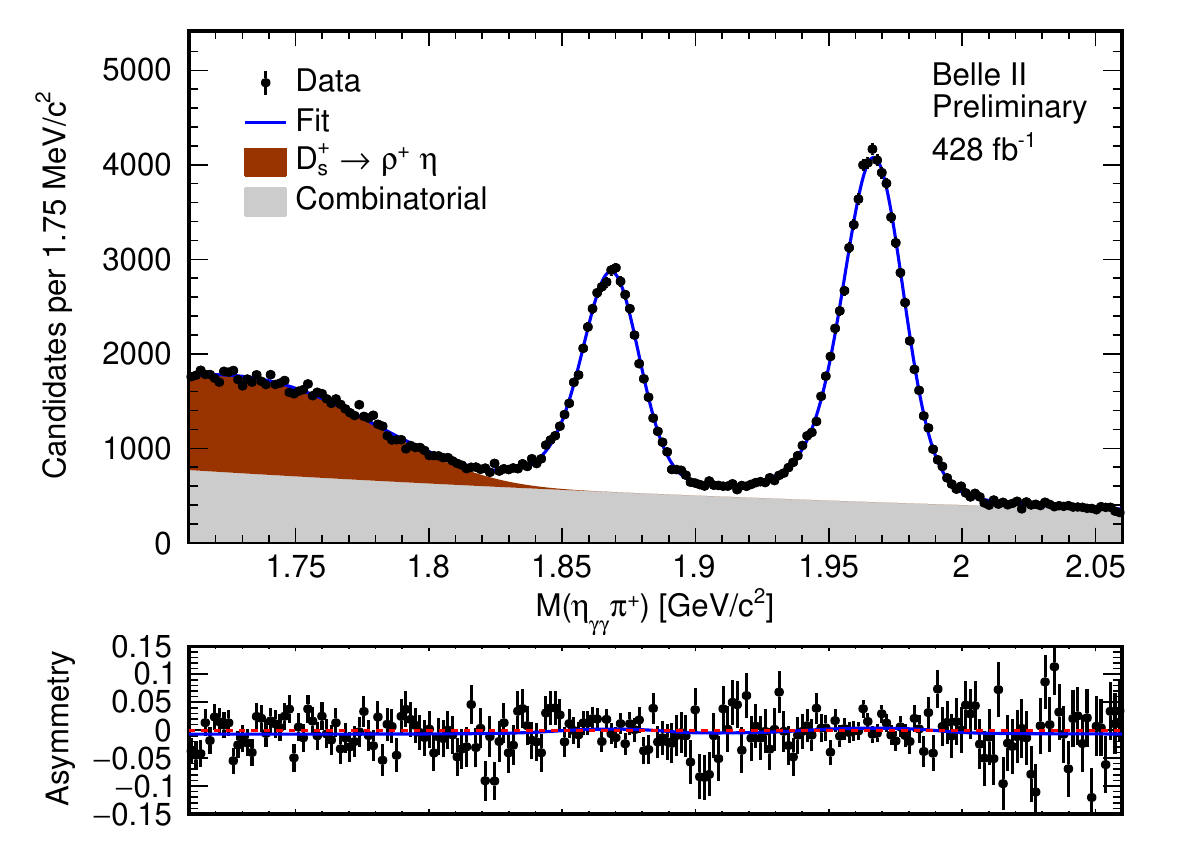}\\
  {\small (a) $\eta\to\gamma\gamma$, $h=\pi$}
\end{minipage}\hfill
\begin{minipage}{0.46\linewidth}
  \centering
  \includegraphics[width=\linewidth]{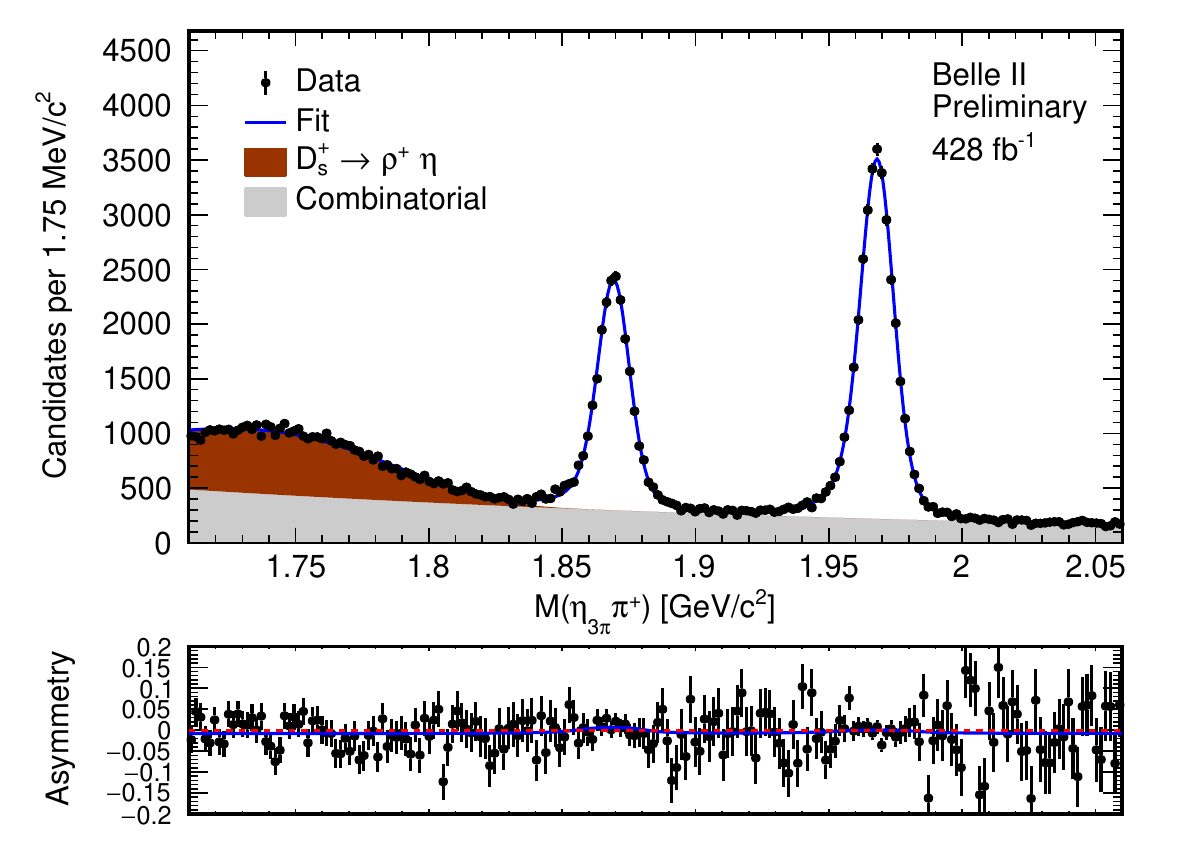}\\
  {\small (b) $\eta\to\pi^+\pi^-\pi^0$, $h=\pi$}
\end{minipage}
\par\medskip
\begin{minipage}{0.46\linewidth}
  \centering
  \includegraphics[width=\linewidth]{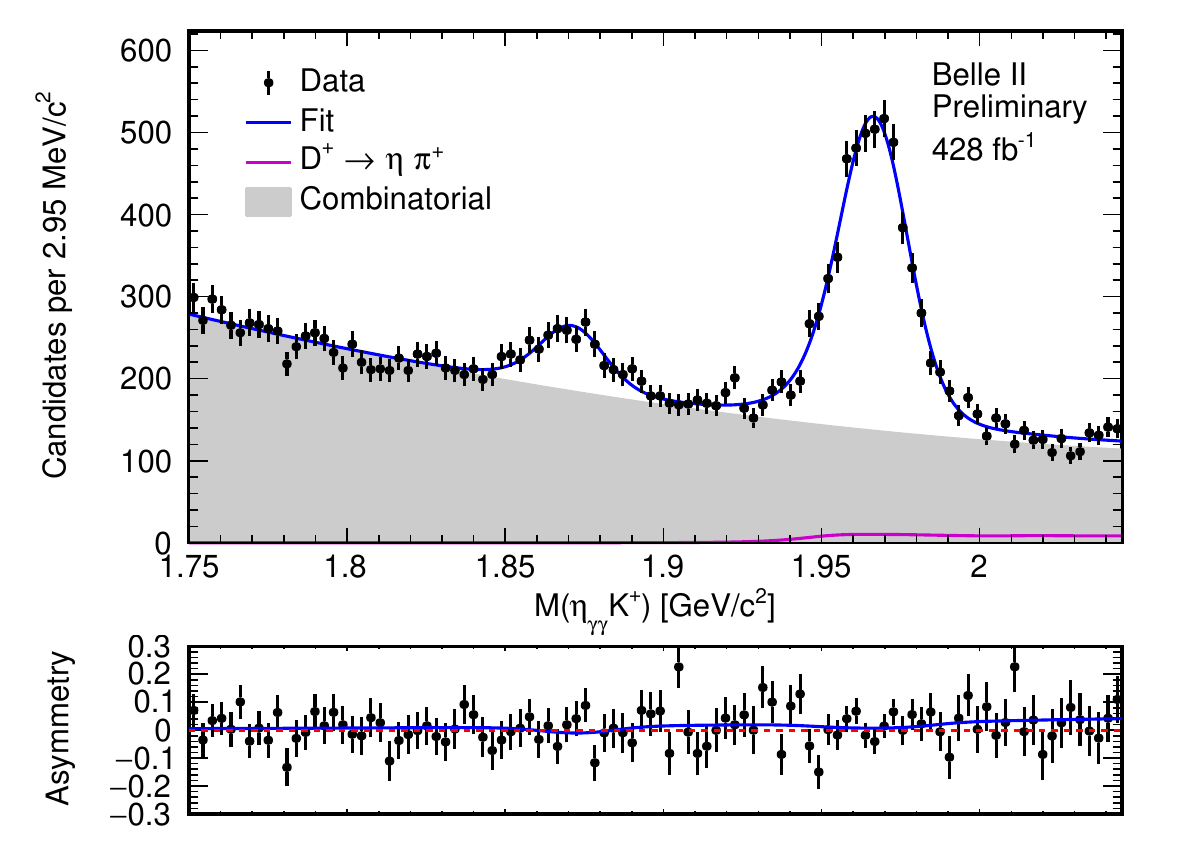}\\
  {\small (c) $\eta\to\gamma\gamma$, $h=K$}
\end{minipage}\hfill
\begin{minipage}{0.46\linewidth}
  \centering
  \includegraphics[width=\linewidth]{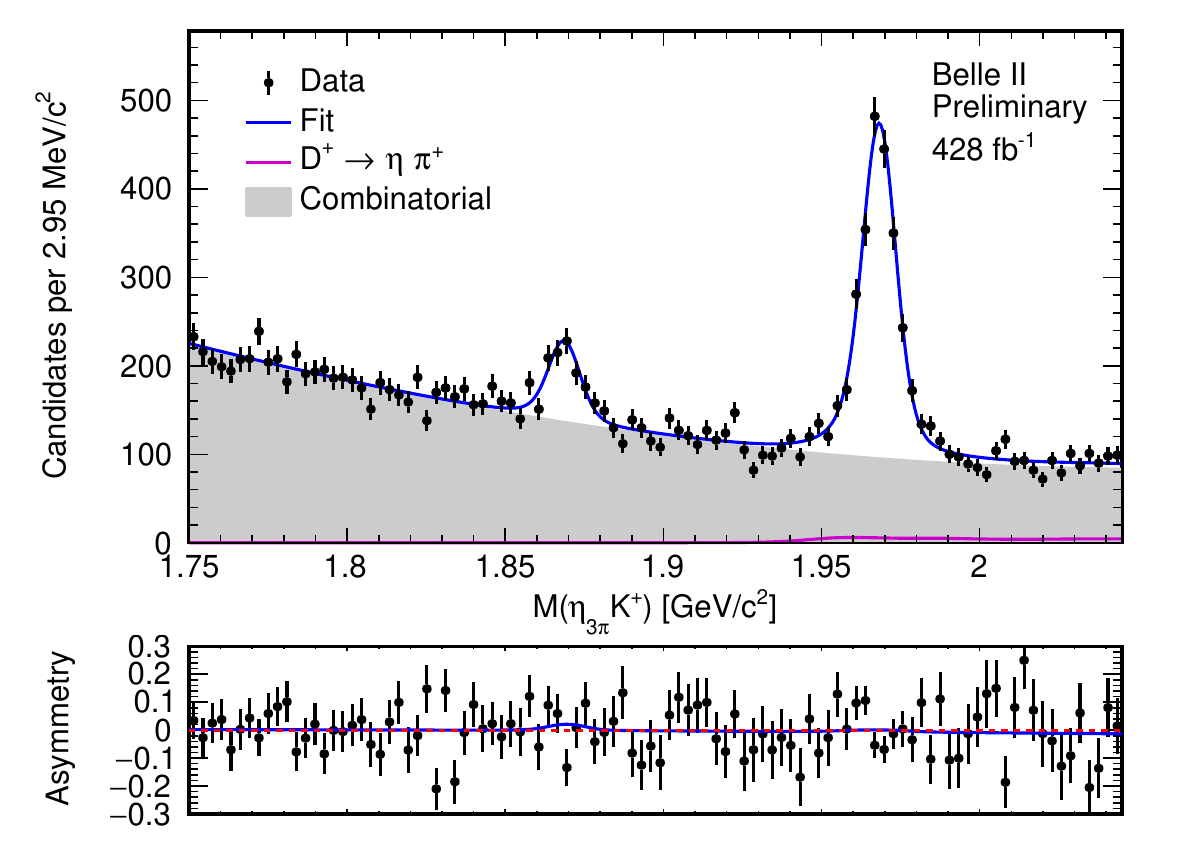}\\
  {\small (d) $\eta\to\pi^+\pi^-\pi^0$, $h=K$}
\end{minipage}
\caption{Preliminary fits to $D^+_{(s)}\to\eta h^+$. Points show data, blue curves total fits, and shaded regions or magenta curves backgrounds. The lower panels show raw yield asymmetries.}
\label{fig:eta-fits}
\end{figure}

Both $\eta\to\gamma\gamma$ and $\eta\to\pi^+\pi^-\pi^0$ are reconstructed and the results are combined.
All $\pi^0$ candidates mentioned hereafter are reconstructed via $\pi^0\to\gamma\gamma$.
Background is highly suppressed using dedicated boosted decision trees (BDTs) for each final state; Figure~\ref{fig:eta-fits} shows the fits.
The CF control modes $D^+\to K_S^0\pi^+$ and $D_s^+\to K_S^0K^+$, for which negligible $C\!P$ violation is expected, are used to correct detection asymmetries.
Table~\ref{tab:eta} lists the preliminary results, consistent with $C\!P$ symmetry.
The $\eta K^+$ measurements are the most precise available as of ICHEP 2026.

\begin{table}[htb]
\centering
\begin{tabular}{lc}
\hline
Decay mode & Preliminary $A_{C\!P}$ (\%) \\
\hline
$D^+\to\eta\pi^+$ & $-0.070\pm0.515\pm0.018$ \\
$D_s^+\to\eta\pi^+$ & $-0.323\pm0.380\pm0.018$ \\
$D^+\to\eta K^+$ & $-1.63\pm6.51\pm0.26$ \\
$D_s^+\to\eta K^+$ & $-0.37\pm1.96\pm0.10$ \\
\hline
\end{tabular}
\caption{Preliminary $C\!P$ asymmetries in $D^+_{(s)}\to\eta h^+$ at Belle II.}
\label{tab:eta}
\end{table}

\subsubsection{\texorpdfstring{$D \to \pi \pi$}{D -> pi pi}}
The $D\to\pi\pi$ system provides complementary tests of the isospin structure of $C\!P$ violation~\cite{isospin}.
In the isospin limit, the $D^+\to\pi^+\pi^0$ final state has $I=2$ and is produced through a $\Delta I=3/2$ transition.
SM QCD penguin contributions instead have $\Delta I=1/2$, making a measurable asymmetry in this charged mode a sensitive test of new physics, subject to small electroweak and isospin-breaking effects. The neutral modes can contain both $I=0$ and $I=2$ amplitudes.

The $D^+\to\pi^+\pi^0$ analysis fits the $D^+$ mass in $D^{*+}\to D^+\pi^0$-tagged and untagged samples, using $D^+\to K_S^0\pi^+$ as a control channel~\cite{dpipi}.
The $D^0\to\pi^0\pi^0$ analysis uses a $D^{*+}$ tag, BDT selection, and a fit to $M(D^0)$ and $\Delta M\equiv M(D^{*+})-M(D^0)$.
Tagged and untagged $D^0\to K^-\pi^+$ controls correct slow-pion detection effects; forward--backward averaging removes production asymmetries~\cite{d00}.
The results are
\begin{align}
A_{C\!P}(D^+\to\pi^+\pi^0)&=(-1.8\pm0.9\pm0.1)\%,\\
A_{C\!P}(D^0\to\pi^0\pi^0)&=(0.30\pm0.72\pm0.20)\%.
\end{align}
Both are consistent with $C\!P$ symmetry.
The charged-mode measurement reduces the statistical uncertainty by approximately $30\%$ relative to the Belle measurement.
The $\pi^0\pi^0$ signal requires efficient photon reconstruction and rejection of beam-background and hadronic clusters.
Its main systematic uncertainties arise from signal and control fit models and from kinematic matching~\cite{d00}.

An isospin test combines the three direct asymmetries.
The corresponding partial widths are $g_{+-}=\mathcal B_{+-}/\tau_{D^0}$, $g_{00}=\mathcal B_{00}/\tau_{D^0}$, and $g_{+0}=\mathcal B_{+0}/\tau_{D^+}$, where $\tau_{D^0}$ and $\tau_{D^+}$ are the neutral-$D$ and charged-$D$ meson lifetimes.
With $a_{ij}=A_{CP}^{\rm dir}(D\to\pi^i\pi^j)$, the normalized combination is
\begin{equation}
R=\frac{g_{+-}a_{+-}+g_{00}a_{00}-\tfrac23g_{+0}a_{+0}}
{g_{+-}+g_{00}-\tfrac23g_{+0}}.
\end{equation}
This combination isolates a $C\!P$-odd contribution from the $\Delta I=1/2$ amplitude, after correcting neutral-mode asymmetries for indirect $C\!P$ violation.
A nonzero $R$ would establish $C\!P$ violation in this isospin sector, complementing the $\Delta I=3/2$ test in $D^+\to\pi^+\pi^0$~\cite{isospin}.
The published $\pi^0\pi^0$ analysis gives $R=(1.5\pm2.5)\times10^{-3}$~\cite{d00}, improving precision by $20\%$.

\subsubsection{\texorpdfstring{$D^0 \to K_S^0 K_S^0$}{D0 -> Ks Ks}}
Interference involving $W$-exchange amplitudes can enhance the SM asymmetry in $D^0\to K_S^0K_S^0$ to the percent level.
The $D^{*+}$-tagged analysis~\cite{ksks} fits $\Delta M$ and the smaller $K_S^0$ flight significance to separate $D^0\to K_S^0\pi^+\pi^-$ background.
The control channel $D^0\to K^+K^-$ corrects production and detection asymmetries, with its known $A_{C\!P}$ taken into account.

The independent CFT analysis~\cite{ksksos} excludes $D^{*+}$-tagged candidates.
A dedicated BDT suppresses background.
The CFT dilution is calibrated separately for Belle and Belle II using $D^0\to K^-\pi^+$ decays.
The calibrated per-candidate dilution enters the likelihood, retaining the information on tag quality.
Fits to the $D^0$ mass and tagger output yield
\begin{align}
A_{C\!P}^{D^*\text{-tagged}}&=(-1.4\pm1.3\pm0.1)\%,\\
A_{C\!P}^{\text{opposite-side}}&=(+1.3\pm2.0\pm0.2)\%,\\
A_{C\!P}^{\text{combined}}&=(-0.6\pm1.1\pm0.1)\%.
\end{align}
The inclusive tag increases the effective statistics by approximately $40\%$; the combined result agrees with $C\!P$ symmetry.

\subsubsection{\texorpdfstring{$D^0 \to \pi^+ \pi^- \pi^0$}{D0 -> pi+ pi- pi0}}
Belle II measures the time- and phase-space-integrated asymmetry in $D^{*+}$-tagged $D^0\to\pi^+\pi^-\pi^0$ decays~\cite{d3pi}. The analysis fits $M(D^0)$ and $\Delta M$ in eight $\cos\theta_{\rm CM}$ bins.
Symmetric-bin averaging and a $D^0\to K^-\pi^+$ control channel cancel production and slow-pion detection effects, yielding
\begin{equation}
A_{C\!P}(D^0\to\pi^+\pi^-\pi^0)=(0.29\pm0.27\pm0.13)\%,
\end{equation}
consistent with $C\!P$ symmetry and $34\%$ more precise than BABAR's result~\cite{d3pi}.
The fit distinguishes correctly tagged signal from genuine $D^0$ decays combined with unrelated tag pions, misidentified $D^0\to K^-\pi^+\pi^0$ decays, and combinatorial background.
Future studies comparing the $D^0$ and $\bar{D}^0$ Dalitz distributions could probe localized $C\!P$ asymmetries that may cancel upon phase-space integration.

\subsection{Charm baryons}
\subsubsection{\texorpdfstring{$\Xi_c^+\to \Sigma^+ h^+ h^-$, $\Lambda_c^+ \to p h^+ h^-$}{Xic -> Sigma h h, Lc -> p h h}}
Belle II reports the first individual decay-rate $C\!P$ asymmetries in these SCS three-body baryon modes, with $h=\pi,K$~\cite{baryon}. The $\Sigma^+$ is reconstructed through $p\pi^0$.
The positive- and negative-charge samples are fitted simultaneously to determine the signal yields, with separate fits for the forward and backward regions.
Forward--backward averaging suppresses production asymmetries.

Detection effects are corrected with kinematically weighted CF control modes, assuming negligible $C\!P$ asymmetries.
The weights match the final-state momentum and polar-angle distributions of the signal and control samples to improve cancellation of detection asymmetries.
For $\Xi_c^+$ measurements, the $\Lambda_c^+\to\Sigma^+h^+h^-$ asymmetry is subtracted.
In the $\Lambda_c^+$ measurements, subtracting the $\Lambda_c^+\to pK^-\pi^+$ asymmetry cancels proton detection effects, and adding the $D^0\to\pi^+K^-\pi^+\pi^-$ asymmetry cancels the introduced kaon--pion detection asymmetry.
The results shown in Table~\ref{tab:baryon} are consistent with $C\!P$ symmetry.
\begin{table}[htb]
\centering
\begin{tabular}{lc}
\hline
Decay mode & $A_{C\!P}$ (\%) \\
\hline
$\Xi_c^+\to\Sigma^+K^+K^-$ & $3.7\pm6.6\pm0.6$ \\
$\Xi_c^+\to\Sigma^+\pi^+\pi^-$ & $9.5\pm6.8\pm0.5$ \\
$\Lambda_c^+\to pK^+K^-$ & $3.9\pm1.7\pm0.7$ \\
$\Lambda_c^+\to p\pi^+\pi^-$ & $0.3\pm1.0\pm0.2$ \\
\hline
\end{tabular}
\caption{$C\!P$ asymmetries in SCS three-body charmed-baryon decays~\cite{baryon}.}
\label{tab:baryon}
\end{table}

These results also test $U$-spin symmetry, which interchanges $d$ and $s$ quarks. In the exact symmetry limit, each of the following sums vanishes~\cite{uspin}; experimentally,
\begin{align}
A_{C\!P}(\Lambda_c^+\to pK^+K^-)+A_{C\!P}(\Xi_c^+\to\Sigma^+\pi^+\pi^-)
&=(13.4\pm7.0\pm0.9)\%,\\
A_{C\!P}(\Lambda_c^+\to p\pi^+\pi^-)+A_{C\!P}(\Xi_c^+\to\Sigma^+K^+K^-)
&=(4.0\pm6.6\pm0.7)\%.
\end{align}
Both sums agree with zero.
The complete $d\leftrightarrow s$ interchange relates $\Lambda_c^+$ and $\Xi_c^+$ decays, including their spectator quarks; it does not give an analogous sum rule between the two $\Lambda_c^+$ modes alone~\cite{uspin}.
These tests reach approximately $7\%$ precision and remain statistically limited. Future tests with larger samples must also account for $U$-spin breaking.

\section{Summary}
All reported $C\!P$ asymmetry measurements are statistically limited and consistent with $C\!P$ symmetry.
The preliminary $\eta h^+$ results probe different Cabibbo transitions, the $D\to\pi\pi$ measurements constrain the isospin structure of $C\!P$ violation, and inclusive flavor tagging increases sensitivity in $D^0\to K_S^0K_S^0$. The first individual three-body baryon asymmetries test $U$-spin relations.
Including Belle II Run~2 data is expected to reduce statistical uncertainties by about $30\%$ for analyses using only Belle II data, improving sensitivity to small asymmetries and strengthening symmetry tests.

\end{document}